\documentclass{elsart}
\usepackage{graphicx}
\usepackage{amssymb}

\newcommand{\loe}{\stackrel{<}{\sim}}
\newcommand{\goe}{\stackrel{>}{\sim}}

\begin{document}
\begin{frontmatter}




\title{Simulation of the AGILE Gamma-Ray Imaging
Detector Performance: Part II}

\author[label1,label4]{Veronica Cocco},
\author[label2,label4]{Francesco Longo},
\author[label3,label1,label4]{Marco Tavani}

\vskip .1in

 \address[label1]{Universit\`a degli Studi ``Tor Vergata'' and INFN
Sezione di Roma II (Italy)}
\address[label2]{Universit\`a degli Studi di Ferrara and INFN Sezione di
Ferrara (Italy)}
\address[label3]{Istituto di Fisica Cosmica, CNR, Milano (Italy) }
\address[label4]{Consorzio Interuniversitario Fisica Spaziale, Torino (Italy)}

\begin{abstract}
In this paper (Paper II)  we complete our discussion on the results
of a comprehensive GEANT simulation
of the  scientific performance of the AGILE Gamma-Ray Imaging
Detector (GRID), operating in the  $\sim 30$~MeV--50~GeV energy range 
in an equatorial orbit of height near 550~km. Here
we focus  on the on-board Level-2 data
processing and  discuss possible alternative strategies for event
selection and their optimization.

We find that the dominant particle background components after our
Level-2 processing are electrons and positrons of kinetic
energies between 10 and 100 MeV penetrating the GRID instrument
from directions almost parallel to the Tracker planes (incidence
angles $\theta \goe 90^{\circ}$) or from below.

The analog (charge) information
available on-board in the GRID Tracker is crucial for a reduction
by almost three orders of magnitude of protons (and heavier ions)
with kinetic energies near 100~MeV.

We also present in this paper the telemetry structure of the GRID
photon and particle events, and obtain the on-board effective area
for photon detection in the energy range $\sim 30$~MeV--50~GeV.

\end{abstract}

\end{frontmatter}


\section{Introduction}

The use of solid state physics instruments for cosmic gamma-ray
detection in space will substantially improve the scientific
performance of high-energy astrophysics missions. AGILE
\cite{agile-2,agile-3} is a Small Scientific Mission supported by
the Italian Space Agency planned to be operational in 2003. AGILE
is a relatively light instrument ($\sim 80$~kg) based on
state-of-the-art Silicon detector technology with excellent
imaging capabilities in the gamma-ray (30~MeV--50GeV) and hard
X-ray (10--40~keV) energy ranges. The Gamma-Ray Imaging Detector
(GRID) is devoted to optimal detection and imaging of cosmic
gamma-rays. It is basically made of a Silicon Tracker and a
Mini-Calorimeter as described in Ref.~\cite{agile-2}. The Silicon
Tracker has 14 planes of Si microstrip detectors (121~$\mu$m
pitch) with floating strip readout (readout pitch of 242~$\mu$m)
for a total on-axis radiation length of 1~$X_0$. The
Mini-Calorimeter, with a total on-axis radiation length of 1.5~$X_0$, 
supports the event energy
determination and topological reconstruction of gamma-ray events.

In this paper we complete our analysis of the on-board data
processing of cosmic photon and charged particle background
events by what we define ``Level-2/Step-1 data processing''. We
refer to a companion paper (Paper I, Longo, Cocco \& Tavani,
2001) for details on the AGILE-GRID model and
assumptions about the background and detector performance
capabilities.

\subsection{Summary of the GRID Level-1 data processing}

As shown in Paper I the best Level-1 trigger strategy (required
to be fast within a few tens of microseconds) is given by a
combination of what we defined as the R11G and the DIS options.
The R11G option is based on the combined use of signals from the
anticoincidence (AC) panels and of the quantity $R$, defined as
the ratio between the total number of hit TAA1 chips and the total
number of fired X and Y views. The DIS option is a simplified
track  reconstruction based on computing the distance $D$ of the
    fired TAA1s  from the fired AC lateral panels.

From our simulations we showed that the R11G+DIS Level-1 trigger
is quite efficient in rejecting $\sim 96\%$ of background charged
particles without affecting significantly the cosmic gamma-ray
detection \cite{paper-1}. Taking into account also the Earth
albedo-photons, we expect a total (background plus cosmic photons)
rate of
 $\loe 100$ Hz after the hardware-implemented Level-1 trigger 
(see Paper I). This rate is consistent with the AGILE Data Handling
(DH) processing requirements.

\subsection{GRID Level-2 data processing}

The two types of background events passing the Level-1 data processing 
are:
\begin{itemize}
\item[1)] charged particle events due to primary and albedo protons,
electrons and positrons (average rate $\sim 70 \, \rm s^{-1}$);
\item[2)] Earth albedo-photon events (average rate
    $\sim 20-40 \, \rm s^{-1}$, depending on the instrument inclination
    with respect to the Earth surface).
\end{itemize}
After the Level-1 data processing, an additional event reduction
is necessary on board to satisfy the GRID telemetry conditions.
The goal is to achieve an event rate (comprehensive of cosmic
gamma-rays and background events) of $\loe 30 \, s^{-1}$. This
Level-2 data processing and its implications are the main subjects
of the current paper.

We distinguish two steps of Level-2 processing:
\begin{itemize}
\item  Step-1: simple algorithms using cluster identification,
analog information, and topology of events in the GRID Silicon
Tracker (crucial for particle background rejection);
\item  Step-2: 3D-reconstruction algorithms
 aimed at determining
the incoming photon directions (crucial for rejecting Earth albedo-photons).
\end{itemize}
In the following, we present the main results of the simulated
charged particle background processing (Level-2/Step-1), and the 
requirements for the on-board Level-2 software to be applied to 
albedo-photons.

\section{Level-2 Processing: Step-1}

The Level-2 processing logic is applied after the Level-1 and
Level-1.5 steps, and after the GRID data pre-processing,
consisting in cluster identification and temporary storage in a
GRID memory buffer. The Level-2 processing is asynchronous with
respect to the real GRID data acquisition, and is typically
limited to be completed within 1-2 milliseconds given the GRID
background requirements.
We assume R11G and DIS respectively as the Level-1 and Level-1.5
trigger steps; we define as ``cluster'' every group of
consecutive fired Silicon strips with energy deposition larger
than 27 keV (corresponding to 1/4 MIP\footnote{MIP means Minimum
Ionizing Particle energy release}), and for every cluster we
assume to have available from the on-board data processing the
centroid (charge-barycentric) positions, cluster widths, and
cluster total charges.
An important factor to consider is the saturation of the GRID
Silicon strip channels. When the energy release is larger than
5~MIP, the charge information saturates to its maximum value.
Complete analog energy information is  then available only for
non-saturated strips, and we correctly simulate this hardware
behavior.

Before discussing some Level-2 processing procedures,
we recall the meaning of some quantities and Level-1 trigger 
steps defined in Paper~I:
\begin{itemize}
\item [] {\bf TRA} =  number of events
characterized by primary particles or photons reaching
 the Tracker volume, a box of
$38.06\times 38.06\times 21.078$ $\rm cm^{3}$ which includes the
Tracker planes from the top sheet of the first tungsten layer to
the bottom sheet of the last Silicon-y plane;
\item [] {\bf PLA} = events which give hits in at least
    3 out of 4 consecutive planes (X OR Y view);
\item[]{\bf LAT} = events passing the top-AC veto,
    with signals in 0 or 1 lateral AC panels,
    in 2 consecutive AC panels or in 2 AC panels on the same side;
\item[]{\bf R11G} = LAT events with signals in 0 lateral AC panels,
    and LAT events with signals in 1 or 2 AC panels and $R>1.1$;
\item[]{\bf DIS} = simplified track
    reconstruction  based on computing the distance $D$ of the
    fired TAA1s  from the fired AC lateral panel.
    The parameter DIS is defined as: $DIS=D_{fp}-D_{lp}$ where 
    $D_{fp}$ is the distance of the closest
    fired TAA1 to the fired AC lateral panel in the first
    plane, while $D_{lp}$ is the distance of the closest
    fired TAA1 to the fired AC lateral panel in the last plane.
    We require $DIS\geq 0$ for good events. This option is applied 
    only if there are fired AC lateral panels.
\end{itemize}

We discuss here four Level-2 processing procedures, some of them
inspired by the corresponding Level-1 or Level-1.5 trigger
options that successfully reject background particles without
loosing too many cosmic gamma-ray photons:
\begin{itemize}
    \item[1)]{\bf 3PL:}\\
    is a condition more stringent than the PLA defined in Paper~I;
    it requires hits on 3 consecutive planes (X AND Y views).
    \item[2)] {\bf CDIS:}\\
    is the application of the DIS algorithm to clusters instead
    of TAA1 chips. It is based on computing the distance $CD$ of the clusters
    from the single fired AC panel. The parameter $CDIS$ is defined as
    $CDIS=CD_{firstplane}-CD_{lastplane}$ in order to have $CDIS\geq
    0$ for good events (in case of a plane with more than one cluster,
    it is considered only the nearest cluster to the fired AC panel).
    This option is applied only if there are fired AC lateral panels.
    \item[3)] {\bf FCN3MIP:}\\
    this procedure is based on the use of the parameter
     $FCN=N_{c}(E>3\: MIP)/N_{ctot}$
    which is the fractional number of clusters with total energy
    larger than 3 MIPs ($N_{c}$), with $N_{ctot}$ the total cluster
    number for the whole event; all events with $FCN>0.6$ are rejected.
    \item[4)] {\bf M15}:\\
    The multiplicity M is the analogous of the ratio R, computed for
    clusters: \\
    M=(total number of clusters)/(total number of interested x/y
    views)\\
    The "M15 procedure" consists in rejecting all events with fired AC
    panels and with $M<1.5$.
\end{itemize}

\subsection{Simulation results and discussion}
\begin{figure}[!hb]
    \centering
    \includegraphics[width=\textwidth]{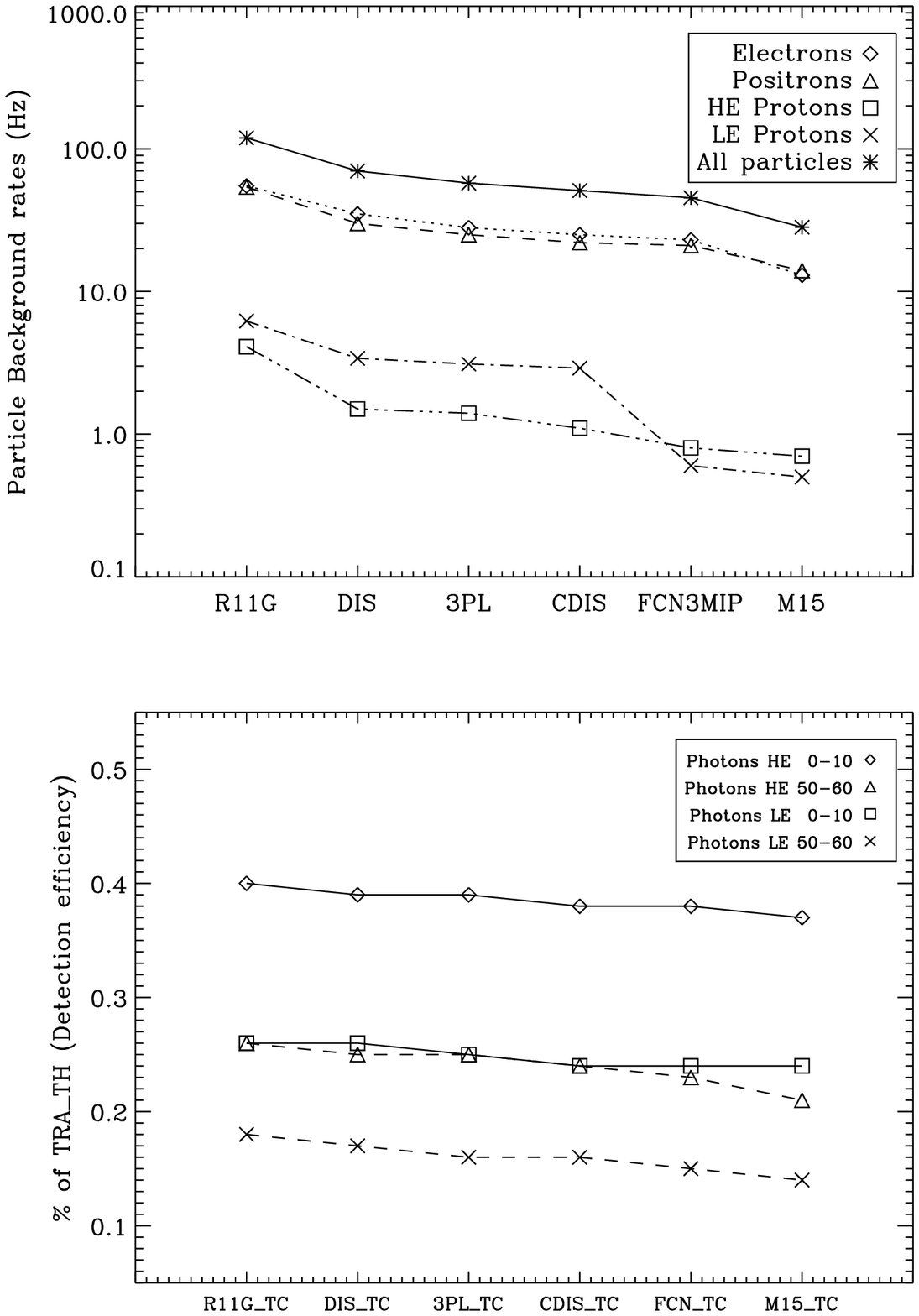}
\caption{\label{fig:trig2}{\small }}
\end{figure}

Tabs.~\ref{tab:trig2_bkg}, \ref{tab:trig2_alb}, \ref{tab:trig2_fot}
  and Fig.~\ref{fig:trig2} show the
simulation results obtained applying the 3PL, CDIS, FCN3MIP and
M15 procedures as Level-2 data processing steps applied in
sequence.
The particle and photon classes used in the simulations are the
same used in Paper I. The
suffix ``TC'' means ``Tracker converted'': only photons converted
in the Tracker volume are ``good photons'', those for which there
is good probability to reconstruct the incident direction.

We note that the 3PL procedure rejects events which are difficult
to interpret, because of their "sparse" topology. This kind of
events are typically produced by background electrons or
positrons rather than by cosmic photons.
The CDIS and M15 procedures follow the same philosophy of
the Level-1.5 DIS option and of the Level-1 R11G option. The main
difference is that they are applied to clusters instead of TAA1
chips, and therefore the spatial resolution is clearly better.
The FCN3MIP procedure uses in a crucial way the GRID cluster
analog information, and is very efficient in rejecting low-energy
protons stopping in the Tracker volume. From Fig.~1 (upper panel,
points marked with crosses) we note that the FCN3MIP procedure
rejects low-energy protons by almost an order of magnitude, and
has a very small effect on the rejection of cosmic off-axis
gamma-rays (Fig.~1, lower panel). This is one of the most
important results of our paper. Ionization losses of protons (or
heavier nuclei) decelerating within the Tracker and eventually
stopping inside it leave an unambiguous signature in terms of
deposited charge in the Silicon microstrips. Our results on the
proton background rejection are also clearly shown in
Fig.~\ref{fig:spec_prot}, indicating a suppression by nearly one
order of magnitude of the surviving flux from Level-1 to
Level-2/Step-1 near kinetic energies of 100~MeV. At these
energies, the total proton background suppression obtained
on-board is by three orders of magnitude, by far the best results
obtained by our background subtraction procedures.
 Our understanding of
the particle background for an equatorial orbit of height near
550~km indicates that protons (and heavier nuclei) contribute
about 10--20\% of the total background rate of incident particles.
An efficient rejection of this component is therefore very
important.

We can conclude that simple Level-2 processing strategies can
succeed in lowering the particle  background rate from $70\;
s^{-1}$ to $\sim 30 \; s^{-1}$  without affecting significantly
the cosmic gamma-ray detection.

\subsection{Spectral and angular selection of background components}

\begin{figure}[!h]
    \centering
    \includegraphics[width=\textwidth]{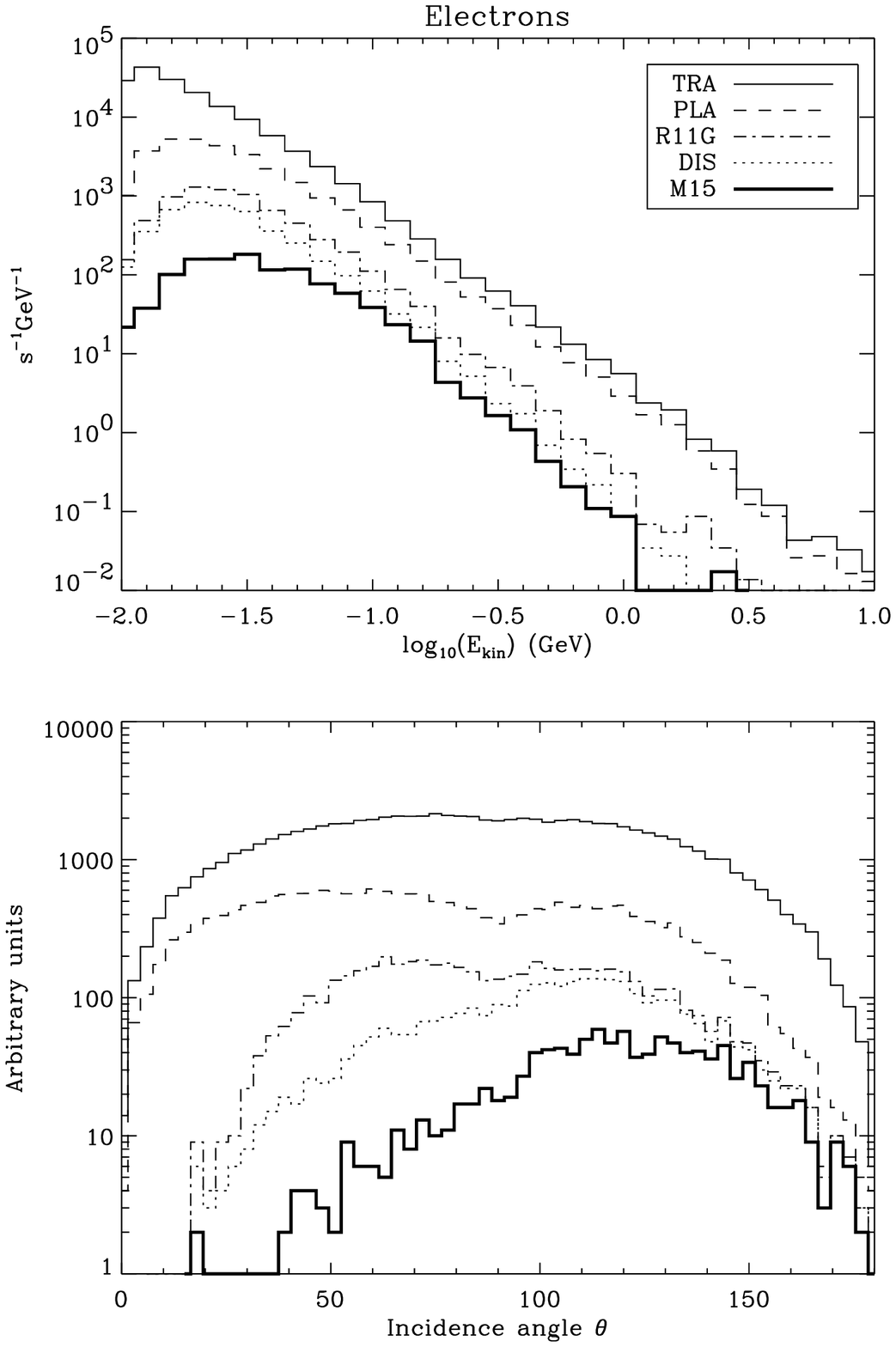}
\caption{\label{fig:spec_ele}}
\end{figure}

\begin{figure}[!h]
    \centering
    \includegraphics[width=\textwidth]{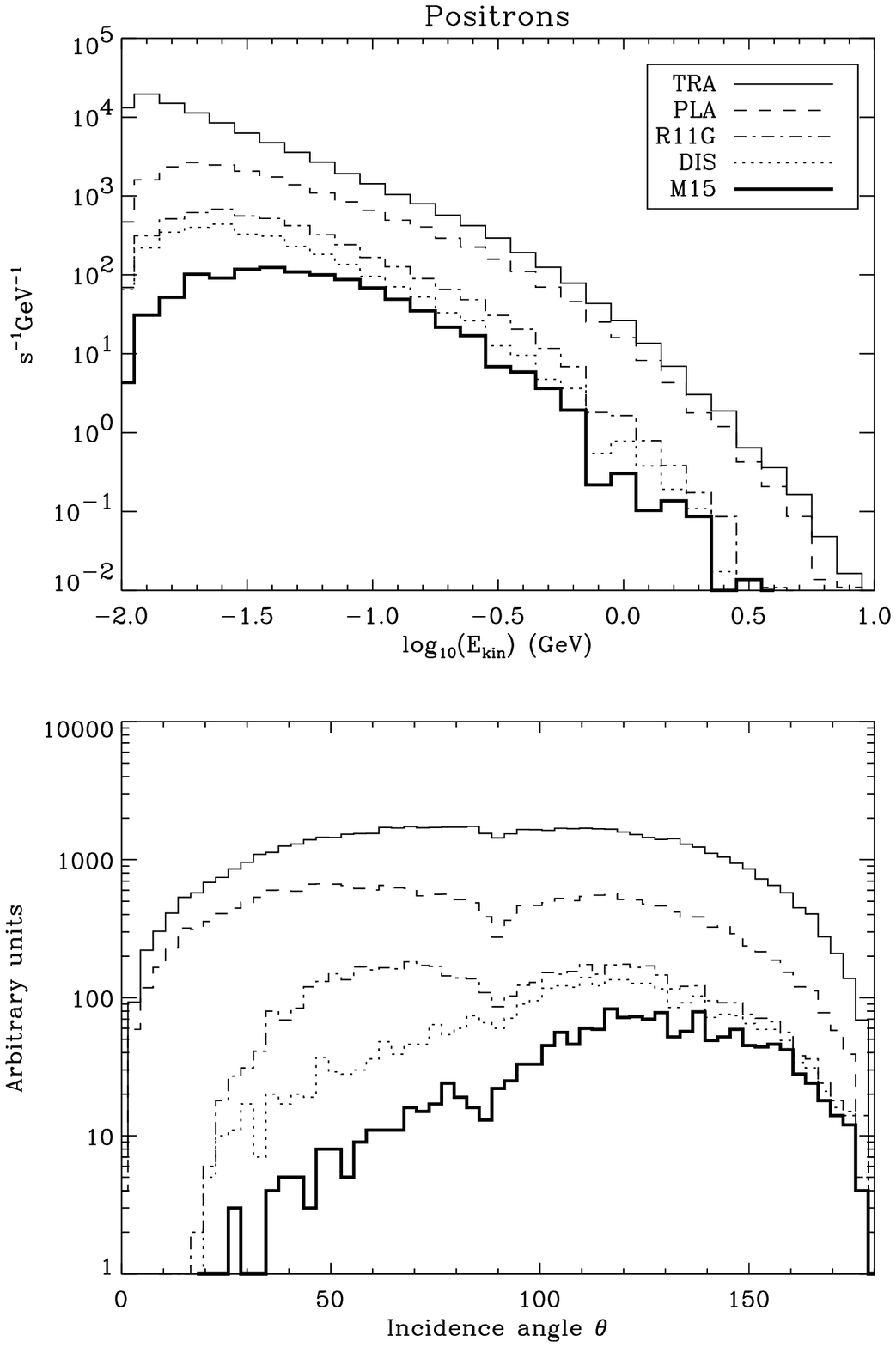}
\caption{\label{fig:spec_pos}}
\end{figure}

\begin{figure}[!h]
    \centering
    \includegraphics[width=\textwidth]{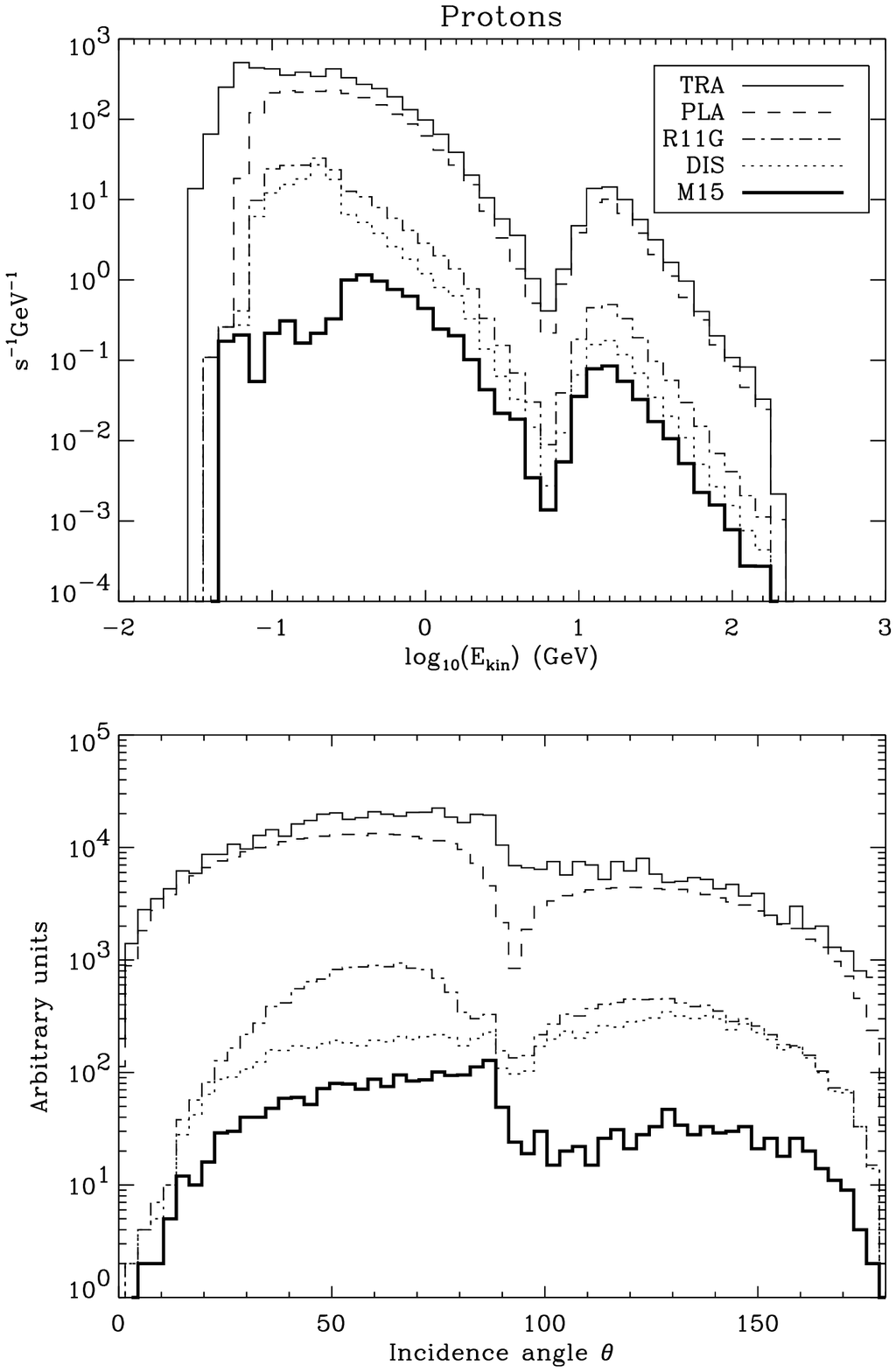}
\caption{\label{fig:spec_prot}}
\end{figure}

We extensively studied  how the different trigger cuts modify the
energy spectra and the angular distributions of the different
background components.
Figs.~\ref{fig:spec_ele}, \ref{fig:spec_pos} and
\ref{fig:spec_prot} show the modifications of the charged
particle background spectra and angular distributions due to
Level-1, Level-1.5 and Level-2 data processing. Note that
low-energy protons are rejected especially by the Level-2/Step-1
trigger selection and that most ``surviving'' particles are
characterized by large values of the incidence angle
($\theta>60^\circ$).
This important qualitative feature of the surviving particle
background applies also to electrons and positrons that
constitute the majority of particles passing the Level-2/Step-1
processing. From Figs.~\ref{fig:spec_ele} and~\ref{fig:spec_pos}
(lower panels) it is evident that particles penetrating in the
GRID from below (with respect to the detector's Z-axis pointed in
a direction opposite to that of the spacecraft) have a larger
probability of passing the Level-2/Step-1 data processing. This
conclusion is not surprising  considering the shallowness of the
Mini-Calorimeter and the existence of lateral GRID regions not
covered by the Anticoincidence panels (Paper~I). It is important
to note that the AGILE-GRID will be an imaging gamma-ray
instrument quite different from EGRET \cite{EGRET} that
could discriminate against particles impinging on the detector
from below because of a Time-of-Flight veto system. Background
reduction for particles penetrating Silicon detectors similar to
AGILE from below is a delicate matter, and needs to be addressed with
great care.  

\begin{figure}[!ht]
    \centering
    \includegraphics[width=\textwidth]{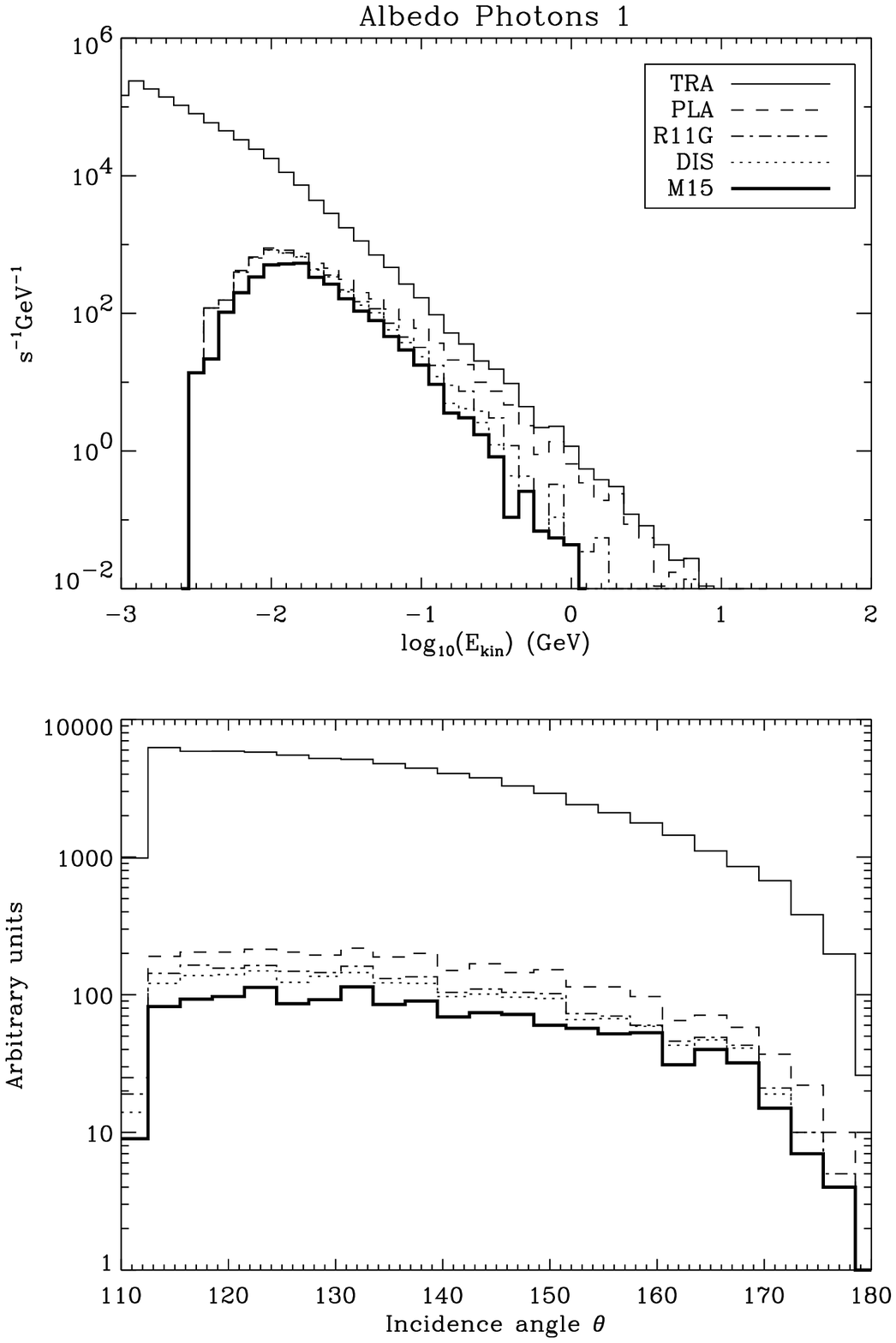}
\caption{\label{fig:spec_alb1}}
\end{figure}

\begin{figure}[!ht]
    \centering
    \includegraphics[width=\textwidth]{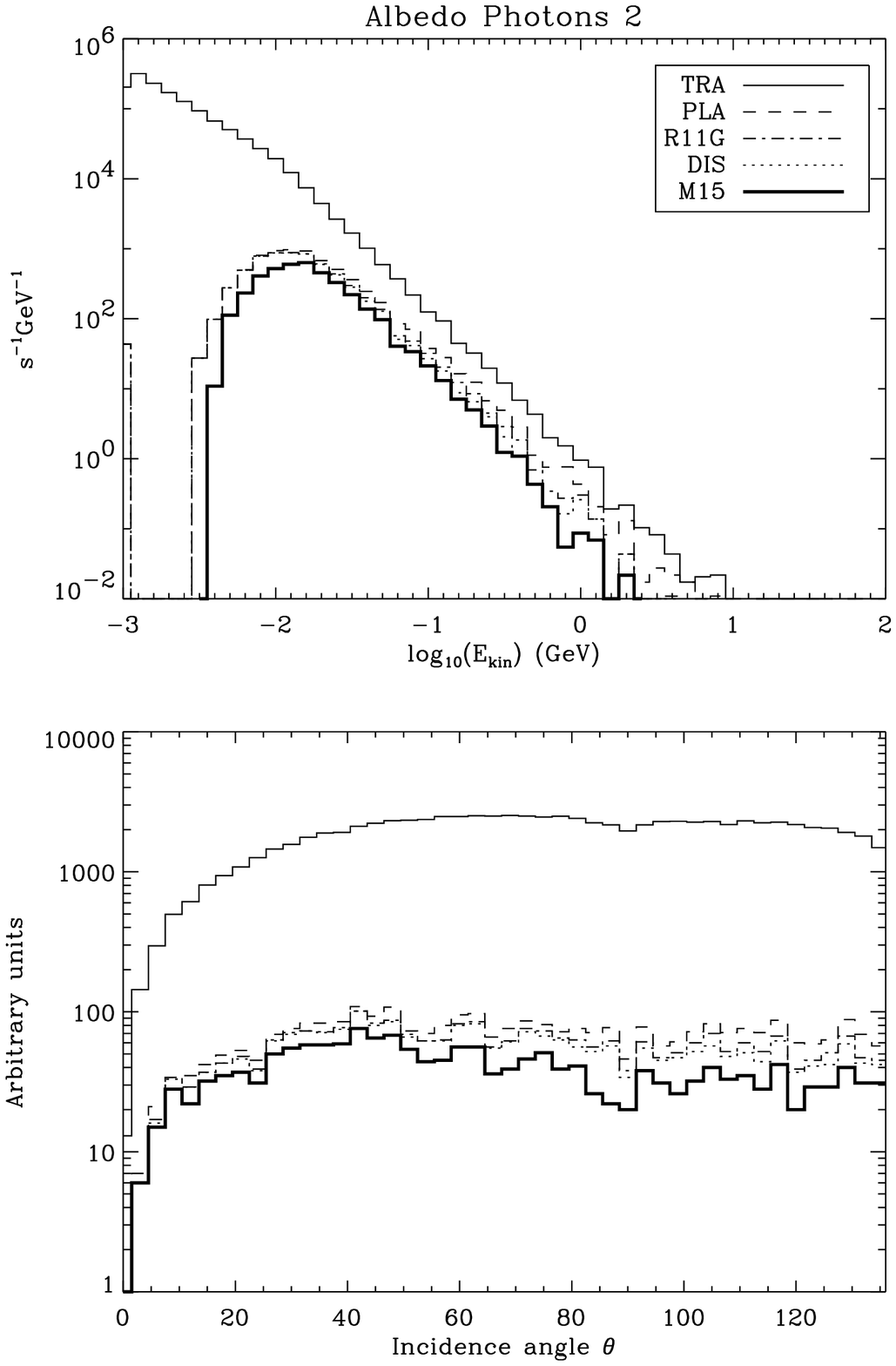}
\caption{\label{fig:spec_alb2}}
\end{figure}

Figs.~\ref{fig:spec_alb1} and~\ref{fig:spec_alb2} show the event
selection and cuts for the albedo-photon spectra and their angular
distributions for different GRID-Earth geometries. We base our
analysis of Earth albedo photons on the 
simplified model described in Paper~I.
 Fig.~\ref{fig:spec_alb1} refers to the
case of the GRID pointing an unocculted portion of the sky with
the Earth "below the GRID" and the direction towards the Earth
center corresponding to the colatitude angle $\theta =
180^{\circ}$. Fig.~\ref{fig:spec_alb2} refers to the case of the
Earth occulting approximately half of the GRID field of view. The
most relevant feature of these albedo gamma-ray events is their
large contribution to the total GRID background after the
Level-2/Step-1 processing. Their differential spectra peak
slightly below $10^3 \;\rm s^{-1} \, GeV^{-1}$ at photon energies
near 10~MeV, and their total rate integrated over the whole
spectrum is relatively high, of the same order as the surviving
lepton rate (see Table~2). This result indicates the necessity of
implementing on board an additional data processing for rejecting
efficiently Earth albedo photons based on their incoming
directions. This analysis goes beyond the scope of this paper and
will be presented elsewhere.

\subsection{\label{par:cphot}Spectral and angular selection of cosmic
gamma-ray photons}

\begin{figure}[!ht]
    \centering
    \includegraphics[width=\textwidth]{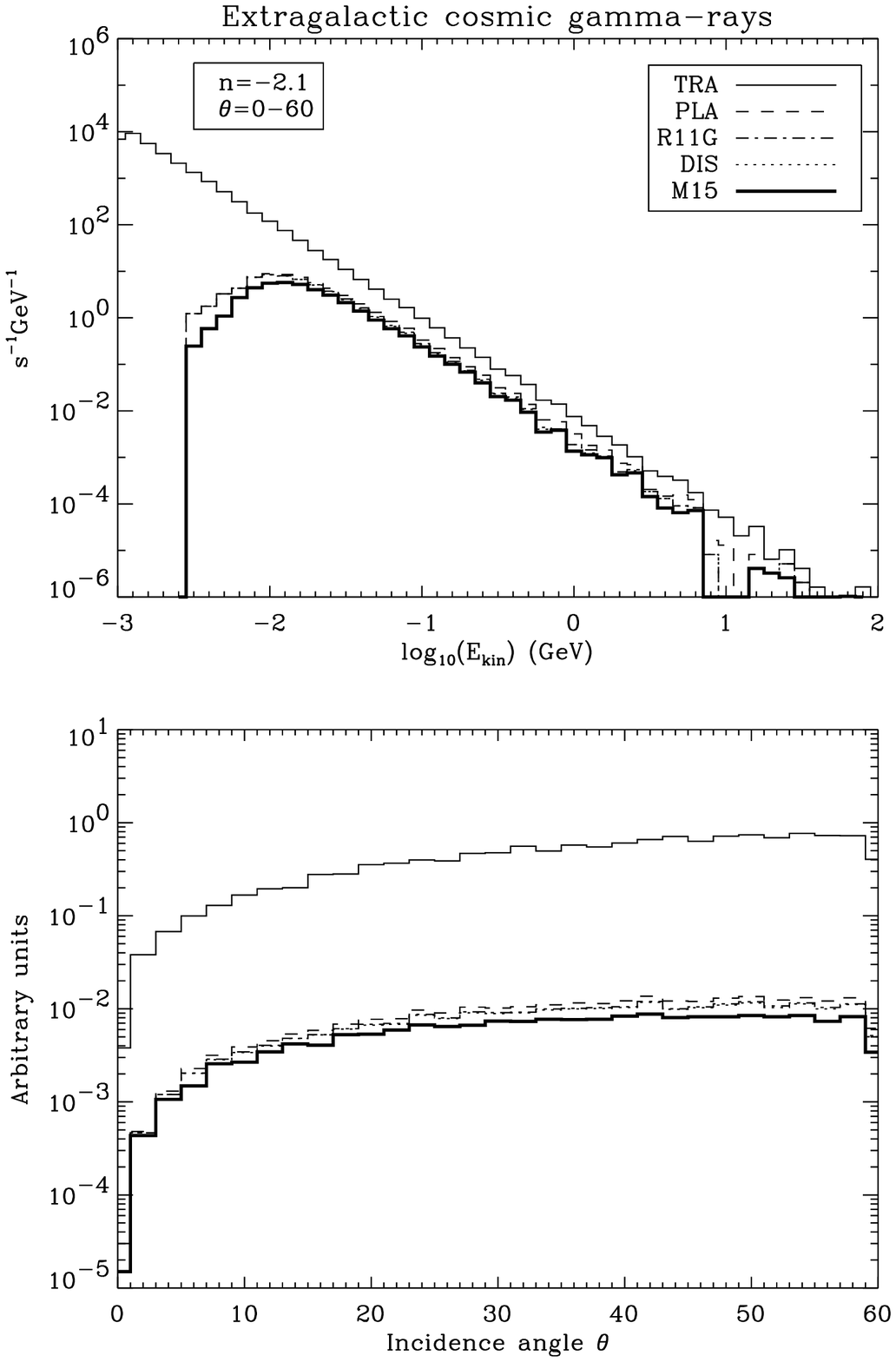}
\caption{\label{fig:spec-eg}}
\end{figure}

In order to analyze the effect of the trigger selection on the
cosmic gamma-ray photon spectrum and angular distribution we considered
extragalactic cosmic gamma-rays
with a power-law energy spectrum of index n=-2.1 and flux
$\Phi \rm (E > 100\, MeV) \simeq 10^{-5}
\; ph \, cm^{-2} \, s^{-1} \, sr^{-1}$ from Ref.~\cite{extragal},
energies in the range 1 MeV $\div$ 100 GeV
and directions in the ranges $\theta=0^{\circ}-180^{\circ}$,
$\phi=0^{\circ}-360^{\circ}$.
Fig.~\ref{fig:spec-eg} shows the
effects of trigger and processing
cuts on the cosmic gamma-ray spectrum and angular distribution.
We notice the excellent trigger performance of the GRID  in terms
of both spectral and angular responses. Trigger efficiency for
photon detection  and Level-2 successful processing varies between
15\% and $\sim$ 40\% depending on photon energy and direction.

\section{Level-2 Processing: Step-2 and software requirements}

The Earth albedo-photon component of the background is of great
relevance.
After the simplified processing of Level-1 and Level-2/Step-1, we
can state the following:
\begin{itemize}
\item[a)] the albedo photon background after Level-2/Step-1 is
dominated by low-energy photons in the range $\sim 5 {\rm MeV} < E < 30$~MeV,
peaking at $\sim 10$~MeV.
\item[ b)] the Level-2/Step-1 albedo photon event rate is near 20--30~$s^{-1}$ and,
when summed with the charged particle net rate, is too large to be
sustained by the AGILE telemetry.
\end{itemize}

Therefore, the
on-board background suppression requires further software  data processing
after the "simplified" Step-1 analysis presented in the previous section.
 We call this processing "Level-2/Step-2",
aimed at an approximate but effective photon direction
reconstruction. A detailed description of this Level-2 processing
is beyond the scope of this paper, and it will be presented
elsewhere.

\section{\label{par:telem}GRID Telemetry}

We summarize in this Section the main characteristics of the GRID
scientific telemetry. Based on the selection cuts operated at
Level-1 and Level-2 processing stages, we are in a position to
assess the contribution to the scientific telemetry for both the
particle and albedo-photon background and the cosmic gamma-ray signal.

It is crucial to realize that the number of bits $N_{bits}$ generated by 
a typical "GRID event" is
variable, depending on the number
of Tracker clusters ($N_{clus}$), fired Mini-Calorimeter bars 
($N_{bars}$) and other
quantities (e.g.: $N_{tplus}$, the number of TAA1 chips exceeding the 
limit of 8 TAA1 for every 2 consecutive views).
In the Montecarlo  simulations we used the formula: 
$N_{bits} \simeq 176+29 \times N_{bars}+57 \times N_{clus}+9 \times
N_{tplus} $, defining 
a ``cluster'' as a  group of consecutive hit
readout-strips with deposited charge $E>1/4 MIP$, and a  
``hit bar'' every CsI bar with an energy
release larger than $E=0.7$ MeV. The considered formula represents the
typical telemetry for GRID events. It takes into account the
event header information and the main contributions of variable
length.
We emphasize that the assumed number of bits per cluster ($n=57$)
includes the total cluster width and deposited charge and all the analog 
information (position and deposited
charge) that can be stored  for 5 readout strips per cluster.

\subsection{\label{par:telclasses}Telemetry event classes}

The relevant components of the expected event rate after the Level-1 and the
Level-2 trigger stages were simulated using the following event classes:

\begin{itemize}
\item[(A)] {\bf Electrons and positrons} (isotropic distributions), this class
includes electron and positron classes described in Paper I;
\item[(B)] {\bf Protons} (including primary and secondary components with
 proper angular distributions, AGILE pointing assumed to be with zenith angle
$\theta = 0^\circ$), this class
includes low-energy proton and high-energy proton classes described in 
Paper I;
\item[(C)] {\bf Earth albedo photons } (case ALB-1, unocculted AGILE's FOV,
Earth below the Tracker), this class is the same considered in
Paper I;
\item[(D)] {\bf Cosmic gamma-rays} (extragalactic diffuse emission),
this class was described in Sect.~\ref{par:cphot} of this paper.
\end{itemize}
Since the average number of bits per GRID event strongly
depends on the particle/photon energy and inclination
and since gamma-rays above hundreds of MeV constitute a very
important component of the scientific data, we considered also
the high-energy and very high-energy photon classes summarized
in Tab.~\ref{tab:hephot}.

\subsection{Simulation results}

Simulation results are summarized in Tab.~\ref{tab:telem1}.
We find that  the lepton component of the background is expected to
 dominate the GRID scientific telemetry. 
We note that the lepton surviving the
current Level-2 cuts are dominated by low-energy events ($E \sim
20-30$~MeV) with characteristics similar to those of cosmic
low-energy gamma-rays. The typical telemetry load for
these low-energy leptons is below 1.5 kbit/event.
Low-energy protons are efficiently rejected by the Level-1 and
Level-2/Step-1 logic.
Note that the telemetry distributions of photon classes are
biased towards the low-energy photons. The average number of bits
per GRID event strongly depends on the photon energy and
inclination.

\begin{figure}[!ht]
\begin{center}
\includegraphics[width=\linewidth]{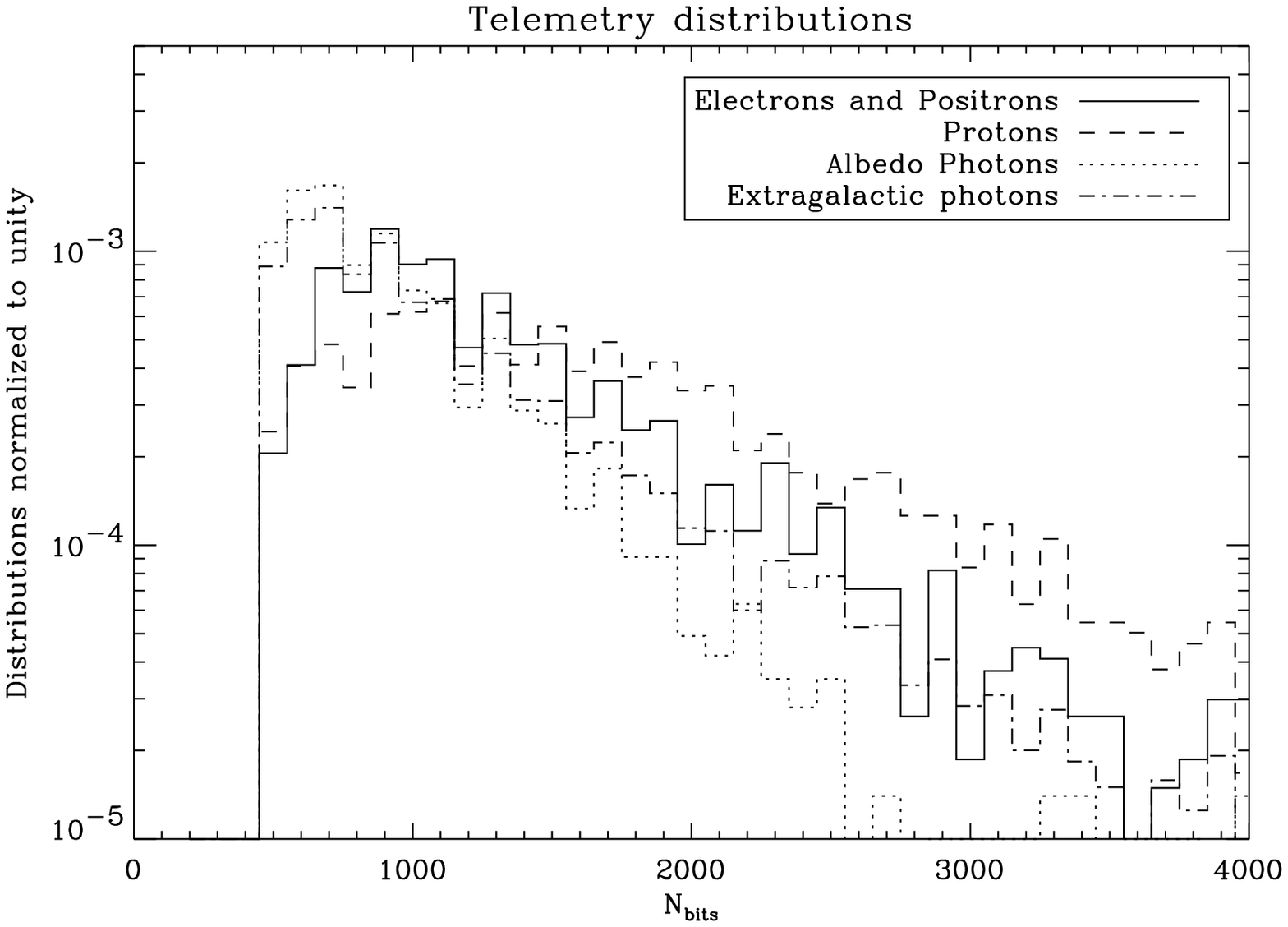}
\caption{\label{fig:telgen}}
\end{center}
\end{figure}

In principle, each particle and photon event is characterized by
different GRID topologies, and therefore different telemetry
loads. However, in practice all particle/photon components
passing the Level-2 processing have quite similar $N_{bits}$
distributions, as shown by Fig.~\ref{fig:telgen}. All
distributions peak near or below 1~kbit/event with average
numbers given in Tab.~\ref{tab:telem1}.

\section{GRID Effective Area}

The effective area is, by definition,
$ A_{eff}=\epsilon A_{\perp}$,
where $ A_{\perp}$ is the detector ``geometrical area''
(equivalent area perpendicular to the incident flux direction) and
$\epsilon$ is the detector efficiency. The detector efficiency is
given by the photon
interaction probability ($\epsilon_{i}$) times the trigger efficiency
($\epsilon_{t}$) times the track reconstruction efficiency
($\epsilon_{r}$):
$ \epsilon=\epsilon_{i}\cdot\epsilon_{t}\cdot\epsilon_{r}$.

Track reconstruction (implying a reliable vertex identification and
direction reconstruction) is strongly influenced by the event
topologies. Taking into account average properties of the events, we
assumed in Fig.~\ref{fig:aeff2}
the following values:
$   \epsilon_r = 1$ for $ E > 25$~MeV and
$   \epsilon_r = 0.75$ for $ E=25$~MeV.

The value of $\epsilon_{i}\cdot\epsilon_{t}$ is given by the following ratio:
$\epsilon_{i}\cdot\epsilon_{t}=\mbox{N(M15\_TC)}/{\mbox{N(TRA\_TH)}}$
since ``good photons'' are only the ones that pass the Level-2 trigger
(R11G+DIS +M15)
having converted in the tracker volume, and they must be compared
with the total number of photons that would geometrically enter the
tracker volume.
N(TRA\_TH) can be evaluated theoretically as
$\mbox{N(TRA\_TH)}=F\cdot A_{\perp}$,
where F is the photon incident flux, which is related to the total
number of events generated on the spherical surface around the
detector by the relation
$F=N_{TOT}/(\pi\,r^{2})$,
where $r$ is the sphere radius (we generally use $r$=89 cm and
$N_{TOT}=50000$).

\begin{figure}[!ht]
\begin{center}
\includegraphics[width=\linewidth]{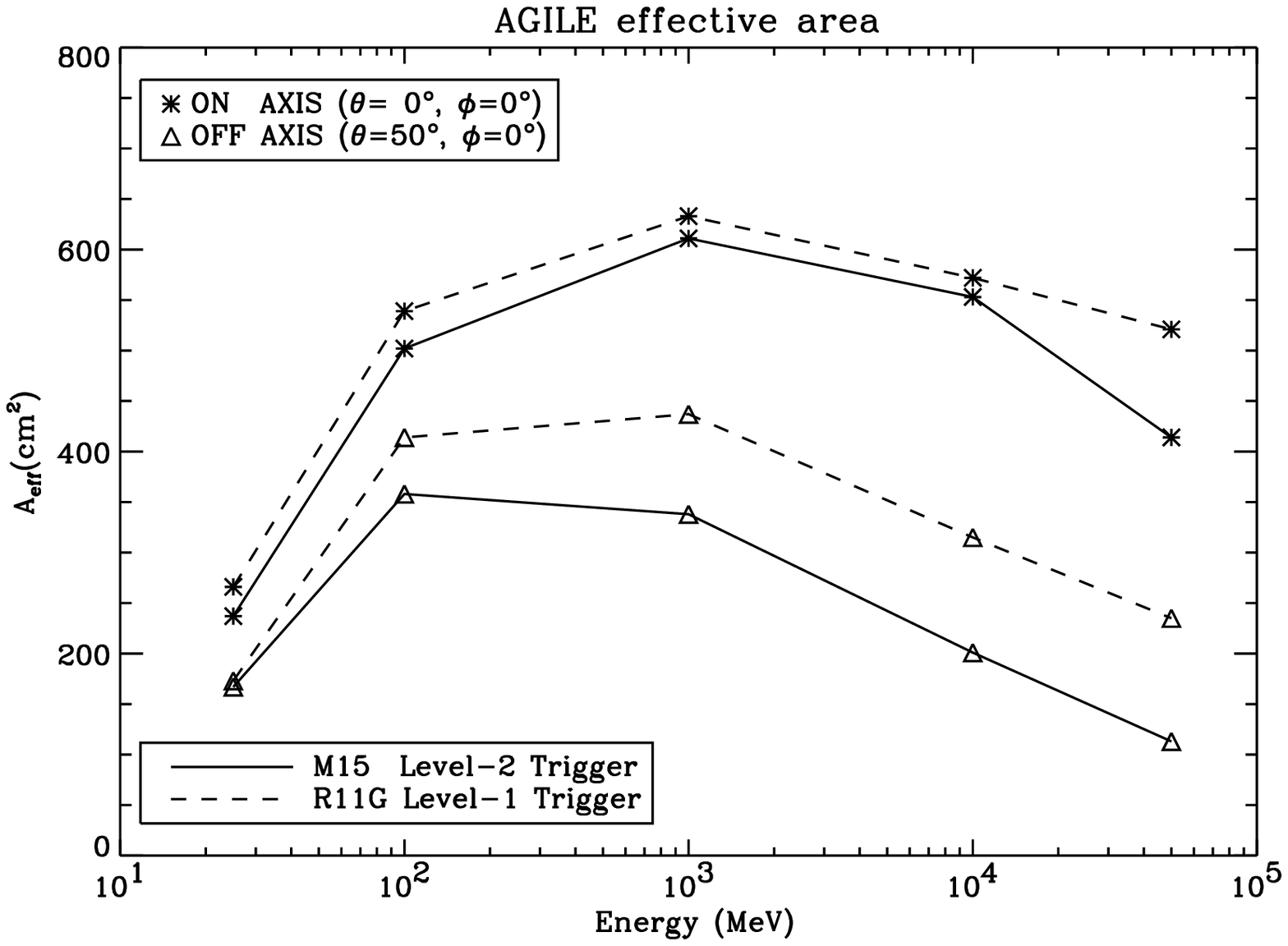}
\end{center}
\caption{\label{fig:aeffcomp}}
\end{figure}

\begin{figure}[!ht]
\includegraphics[width=0.7\linewidth,angle=90]{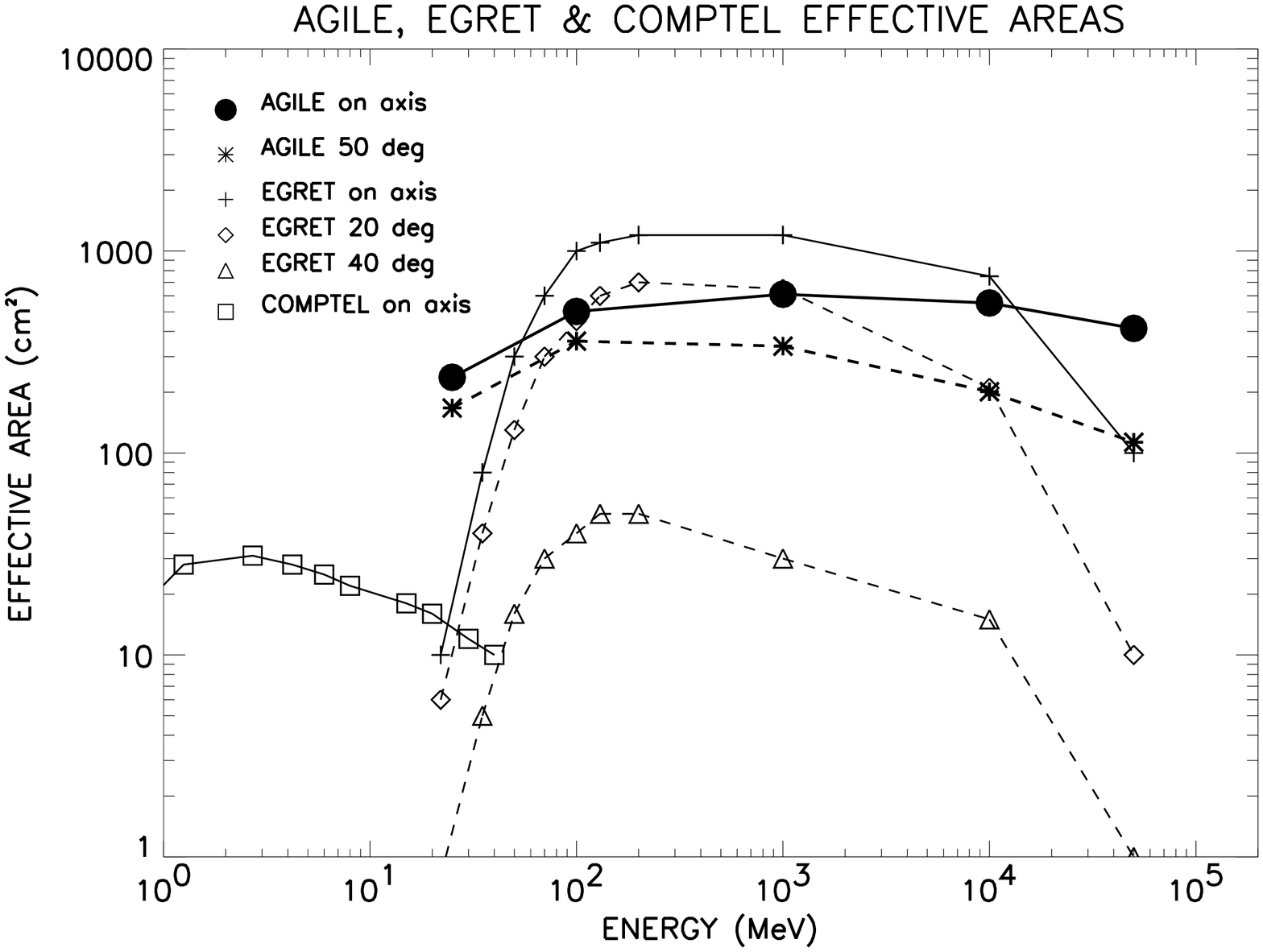}
\caption{\label{fig:aeff2}}
\end{figure}

In order to study the GRID  effective area
we considered on-axis photons ($\theta=0^{\circ}$, $\phi=0^{\circ}$)
and photons with $\theta=50^{\circ}$ and $\phi=0^{\circ}$, with the
following energies: E= 25 MeV, 100 MeV, 1 GeV, 10 GeV, 50 GeV.
Fig.~\ref{fig:aeffcomp} provide information on how the
event cuts adopted in this document affect the GRID gamma-ray detection.
The processing
steps adopted by Level-1.5 and Level-2/Step-1
 are crucial in lowering the particle background rate from
$\sim 120 \, \rm s^{-1}$ (after R11G)
to $\sim 30 \, \rm s^{-1}$ (after Level-2/Step-1).
However,  these event cuts  also cause a decrease of the effective area,
especially for off-axis photons.
Fig.~\ref{fig:aeff2} shows the comparison among AGILE, EGRET and
COMPTEL effective areas, for fixed directions, as a function of photon energy.

\section{Conclusions}

The trigger and processing  strategy  presented in this paper to
filter and select particle and photon GRID events can be
summarized as follows.

Events induced by electrons and positrons constitute the main
background component and dominate the scientific telemetry of the
AGILE-GRID. The total lepton event rate obtained for the trigger
and processing strategy presented in this paper (Level-2/Step-1
processing) is $R_{e^+/e^-} \simeq 30 \; \rm s^{-1}$. A goal of
the Level-2/Step-2 processing through a three-dimensional photon
direction reconstruction is to further reduce this background
component by almost a factor of 2, reaching
$R_{e^+/e^-} \rm (required \; rate) \leq  15 \; \rm s^{-1}$.

Earth albedo gamma-ray photons after the Level-2/Step-1 processing
produce an event rate $R_{albedo-\gamma} \simeq 20-30 \; \rm s^{-1}$
, depending on the geometry and comparable to that of leptons.
 This event rate is too large to be acceptable by
the AGILE telemetry, and further reduction of this component is
necessary. A 3D-direction reconstruction algorithm to be
implemented by the Level-2/Step-2 processing is required to
reduce this rate at least by a factor of 10, reaching the
telemetry rate requirement:
$ R_{albedo-\gamma} \rm (required \; rate) \leq 3 \; \rm s^{-1}$.

Low-energy proton events ($E_{kin} < 400$~MeV) are efficiently
decreased by the on-board Level-1 and Level-2 logic, especially
because of the available Si strip analog information. High energy
protons (of energy near or larger than 1 GeV) tend to dominate the
telemetry of proton events. The simulated Level-2/Step-1
processing of protons produce an event rate  near the required
value, $R_{protons} \rm (required \; rate) \leq 1 \; \rm s^{-1}$.

At the end, the event rate for cosmic gamma-ray events, the scientific signal
of the AGILE-GRID, turns out to be $\sim 100$ times smaller than
the (lepton, proton, albedo-photon) background after the
Level-2/Step-1 processing. As required by our strategy, cosmic
gamma-rays are quite efficiently detected and filtered by the
on-board GRID  data processing, reaching optimal detection
efficiency near 100 MeV.

We notice that a substantial number of cosmic photon events
passing the Level-2 processing have energies between 10~MeV and
30 MeV, as it can be deduced from Fig.~\ref{fig:spec-eg}.
Table~\ref{tab:summa} summarizes our conclusions.

\section{Acknowledgments}

Results presented in this paper are based on joint work with
members of the AGILE Team. In particular, we thank
G.~Barbiellini, P.~Picozza, A.~Morselli and the AGILE Simulation
and Theory Group for many discussions and support.

The current  work was carried out at the  University of Rome "Tor
Vergata", University of Ferrara and CNR and INFN laboratories
under the auspices and partial support of the Agenzia Spaziale
Italiana.

\vspace{1 cm}
{\bf FIGURE CAPTIONS:}\\
\\
Fig.1: Efficiency of the GRID Level-2/Step-1 data processing in rejecting
particle background events (upper panel) and in detecting photons
(lower panel). The suffix ``TC'' means ``Tracker converted''
(only photons converted in the Tracker volume have been
considered) and ``FCN\_TC'' means ``FCN3MIP\_TC'' (see text).\\
Fig.2: Simulated differential energy
(upper panel) and angle (lower panel) distributions resulting
from the processing of the electron background by the AGILE-GRID
on-board Data Handling.  The upper solid curve represents the
particles above 10 MeV penetrating into the Tracker volume (TRA). The
long-dash$\rm ed$ curve
 and the dot-dash$\rm ed$  curves refer
to the Level-1 processing (PLA and R11G), respectively. The short
dash$\rm ed$ curve refers to the Level-1.5 processing (DIS). The
thick solid curve represents the particle flux passing the
sequence of Level-2/Step-1 data processing (indicated with M15).\\
Fig.3: Simulated differential energy
(upper panel) and angle (lower panel) distributions resulting
from the processing of the positron background by the AGILE-GRID
on-board Data Handling.  The upper solid curve represents the
particles  above 10 MeV penetrating into the Tracker volume (TRA). The
long-dash$\rm ed$ curve
 and the dot-dash$\rm ed$  curves refer
to the Level-1 processing (PLA and R11G), respectively. The short
dash$\rm ed$ curve refers to the Level-1.5 processing (DIS). The
thick solid curve represents the particle flux passing the
sequence of Level-2/Step-1 data processing (indicated with M15).\\
Fig.4: Simulated differential energy
(upper panel) and angle (lower panel) distributions resulting
from the processing of the proton background by the AGILE-GRID
on-board Data Handling.  The upper solid curve represents the
particles above 10 MeV penetrating into the Tracker volume (TRA). The
long-dash$\rm ed$ curve
 and the dot-dash$\rm ed$  curves refer
to the Level-1 processing (PLA and R11G), respectively. The short
dash$\rm ed$ curve refers to the Level-1.5 processing (DIS). The
thick solid curve represents the particle flux passing the
sequence of Level-2/Step-1 data processing (indicated with M15).\\
Fig.5: Simulated differential energy
(upper panel) and angle (lower panel) distributions resulting
from the processing of the Earth albedo-photon (ALB-1, see text
for definition) background by the AGILE-GRID. The upper solid
curve represents the photons above 1~MeV penetrating into the
Tracker volume (TRA). The long-dash$\rm ed$ curve
 and the dot-dash$\rm ed$  curves refer
to the Level-1 processing (PLA and R11G), respectively. The short
dash$\rm ed$ curve refers to the Level-1.5 processing (DIS). The
thick solid curve represents the photon flux passing the sequence
of Level-2/Step-1 data processing (indicated with M15).\\
Fig.6: Simulated differential energy
(upper panel) and angle (lower panel) distributions resulting
from the processing of the Earth albedo photon (ALB-2, see text
for definition) background by the AGILE-GRID. The upper solid
curve represents the photons above 1~MeV penetrating into the
Tracker volume (TRA). The long-dash$\rm ed$ curve
 and the dot-dash$\rm ed$  curves refer
to the Level-1 processing (PLA and R11G), respectively. The short
dash$\rm ed$ curve refers to the Level-1.5 processing (DIS). The
thick solid curve represents the photon flux passing the sequence
of Level-2/Step-1 data processing (indicated with M15).\\
Fog.7: Simulated differential energy (upper
panel) and angle (lower panel) distributions from the processing
of cosmic extragalactic gamma-rays by the AGILE-GRID. The upper
solid curve represents the photons above 1~MeV penetrating into
the Tracker volume (TRA). The long-dash$\rm ed$ curve
 and the dot-dash$\rm ed$  curves refer
to the Level-1 processing (PLA and R11G), respectively. The short
dash$\rm ed$ curve refers to the Level-1.5 processing (DIS). The
thick solid curve represents the photon flux passing the sequence
of Level-2/Step-1 data processing (indicated with M15).\\
Fig.8: Telemetry distributions ($N_{bits}$/event) normalized to unity.
{\it Continuous line:} electron-positron component;
{\it Dashed line:} proton component;
{\it Dotted line:} albedo gamma-ray component;
{\it Dashed-dotted line:} extragalactic diffuse gamma-rays.\\
Fig.9: Comparison between the AGILE-tracker effective area obtained
applying only R11G Level-1 Trigger and the AGILE-tracker effective 
area obtained applying M15 Level-2 Trigger (after R11G+DIS).\\
Fig.10: AGILE, EGRET and COMPTEL effective areas after track reconstruction. Figure adapted from
Refs.\cite{agile-2,stefano}. EGRET and COMPTEL data are from
Refs.\cite{galcent,comptel}.

\begin{table}[!ht]
\begin{center}
\caption[10pt]{\label{tab:trig2_bkg}
{\bf Level-2/Step-1 processing effects on Background Charged Particles}}
\vskip .08in
\small
\begin{tabular}{l c c c c c}
\hline
              & ELE & POS  & HE PROT & LE PROT &  TOTAL  \\
&($\rm s^{-1}$)&($\rm s^{-1}$)&($\rm s^{-1}$)&($\rm s^{-1}$)&($\rm s^{-1}$)\\
\hline
R11G          &  55  &  54  &  4.1 & 6.2 & {\bf 119} \\
\hline
DIS           &  35  &  30  &  1.5 & 3.4 &  {\bf 70}  \\
\hline
\hline
3PL           &  28  &  25  &  1.4 & 3.1 &  58  \\
\hline
CDIS          &  25  &  22  &  1.1 & 2.9 &  51  \\
\hline
FCN3MIP       &  23  &  21  &  0.8 & 0.6 &  45  \\
\hline
M15           &  13  &  14  &  0.7 & 0.5 & {\bf 28}  \\
\hline
\end{tabular}
\end{center}
\end{table}
\begin{table}[!hb]
\vspace{1 cm}
\begin{center}
\caption[10pt]{\label{tab:trig2_alb}
{\bf Level-2/Step-1 processing effects on Background Albedo Photons}}
\vskip .08in
\small
\begin{tabular}{l c c}
\hline
              & ALB-1 PHOT & ALB-2 PHOT \\
        &($\rm s^{-1}$)&($\rm s^{-1}$)\\
\hline
R11G          &  22  &  40  \\
\hline
DIS           &  20  &  39   \\
\hline
\hline
3PL           &  16  &  30 \\
\hline
CDIS          &  15  &  29 \\
\hline
FCN3MIP       &  15  &  27 \\
\hline
M15           &  15  &  26   \\
\hline
\end{tabular}
\end{center}
\end{table}
\begin{table}[!h]
\vspace{1 cm}
\begin{center}
\caption[10pt]{\label{tab:trig2_fot}
{\bf Level-2/Step-1 processing effects on Cosmic Gamma-Rays (*)}}
\vskip .08in
\begin{tabular}{l c c c c }
\hline
 Photons             &  HE 0-10 & HE 50-60  &  LE 0-10  & 
              LE 50-60 \\
\hline
R11G\_TC          & 40\% &  26\% &  26\% & 18\%  \\
\hline
DIS\_TC           & 39\% &  25\% &  26\% & 17\%  \\
\hline
3PL\_TC           & 39\% &  25\% &  25\% & 16\% \\
\hline
CDIS\_TC          & 38\% &  24\% &  24\% & 16\% \\
\hline
FCN3MIP\_TC       & 38\% &  23\% &  24\% & 15\% \\
\hline
M15\_TC           & 37\% &  21\% &  24\% & 14\% \\
\hline
\end{tabular}
\end{center}
\vspace{0.3truecm}
(*) We reported the detection efficiencies: the percentages of selected 
events respect 
to the total number of photons that theoretically could enter into the 
Tracker volume (\% of TRA\_TH, as defined in Paper I).
\vspace{0.5truecm}
\end{table}

\begin{table}[!ht]
\begin{center}
\caption{\label{tab:trig2s}
{\bf GRID background rates after Level-2/Step-1 processing}}
\vskip .08in
\begin{tabular}{ l c c}
\hline
Background component        &   unocculted      &   half-occulted \\
            &   GRID FOV        &   GRID FOV    \\
\hline
Charged particles & 30 $s^{-1}$     &    30 $s^{-1}$ \\
Earth albedo photons &  20 $s^{-1}$     &   30 $s^{-1}$  \\
\hline
Total           &   50 $s^{-1}$     &   60 $s^{-1}$ \\
\hline
\end{tabular}
\end{center}
\end{table}
\begin{table}[ht!] \centering \vspace{0.5 cm}
\centerline{\caption{\label{tab:hephot}{\bf Photon classes}}}
\vskip .06in
\begin{tabular}{l r r c c c}
\hline
   Class           & $E_{kin}^{min}$ & $E_{kin}^{MAX}$  & Energy Spectrum &
   $\theta$ & $\phi$ \\
\hline
HE 0-10   &  400 MeV   &  1 GeV  & Power-law (n=-2)  &
$0^{\circ}\div 10^{\circ}$ & $0^{\circ}\div 360^{\circ}$ \\
HE 50-60   &  400 MeV   &  1 GeV  & Power-law (n=-2)  &
$50^{\circ}\div 60^{\circ}$ & $0^{\circ}\div 360^{\circ}$ \\
VHE 0-10   &  1 GeV   &  100 GeV  & Power-law (n=-2)  &
$0^{\circ}\div 10^{\circ}$ & $0^{\circ}\div 360^{\circ}$ \\
VHE 50-60   &  1 GeV   &  100 GeV  & Power-law (n=-2)  &
$50^{\circ}\div 60^{\circ}$ & $0^{\circ}\div 360^{\circ}$ \\
\hline
\end{tabular}
\end{table}
\begin{table}[!ht]
\vspace{0.5 cm}
\begin{center}
\caption{\bf {\small \label{tab:telem1}Average and maximum bit number for different event classes}}
\vskip .06in
\begin{tabular}{ l c c}
\hline
Event class             & Average $N_{bits}$ & Maximum $N_{bits}$ \\
\hline
Electrons/positrons &  1.4 kbit & 5.0 kbit  \\
Protons         &  1.7 kbit & 5.0 kbit  \\
Earth albedo photons    &  1.0 kbit & 3.0 kbit  \\
Cosmic gamma-rays       &  1.1 kbit & 4.0 kbit  \\
\hline
PHOT HE 0-10      & 2.5 kbit     &   6.5 kbit \\
PHOT HE 50-60     & 2.5 kbit     &   7.0 kbit \\
PHOT VHE 0-10     & 2.6 kbit     &   7.0 kbit \\
PHOT VHE 50-60    & 3.1 kbit     &   8.5 kbit \\
\hline
\end{tabular}
\end{center}
\end{table}
\begin{table}[!hb]
\small
\begin{center}
\caption{\bf {\small AGILE-GRID Telemetry Summary}}
\vskip .06in
\begin{tabular}{l c c c r}
\hline
Component &Event Rate ($s^{-1}$)&Rate Req.
($s^{-1}$)&$\frac{<N_{bits}>}{event}$&$<N_T>$/orbit\\
            &(this work)    &(Level-2/Step-2) &   & (1 orbit=5400 sec)\\
\hline
Leptons & 30        & $\leq 20$     &  1.4 kbit    & $\leq$ 151 Mbit \\
Protons         & 1     & $\leq 1 $        & 1.7 kbit  &   $\leq$ ~~~9  Mbit\\
Albedo $\gamma$'s & 20--30  & $\leq 3 $   & 1.0 kbit  & $\leq$ ~~16 Mbit\\
Cosmic $\gamma$'s & 0.1--1 & 0.1--1       & 1.1 kbit  & (0.6--6) Mbit\\
\hline
Total          &         &              &           & 182 Mbit \\
\hline
\end{tabular}
\label{tab:summa}
\end{center}
\end{table}



\addcontentsline{toc}{section}{References}

\begin{thebibliography}{}

\bibitem{agile-2} Tavani, M.,  et al., 2001,
{\em Science with AGILE}, AGILE Internal Note, A-P-019;
http://www.ifctr.mi.cnr.it/Agile.

\bibitem{agile-3} Tavani, M., et al., 2001,
invited paper presented at the {\it Gamma 2001 Symposium},
Baltimore, 4-6 April 2001, to be published by the American
Institute of Physics Conference Proceedings.

\bibitem{paper-1} Longo, F., Cocco, V. \& Tavani, M., 2001,
submitted to NIM (Paper I).

\bibitem{EGRET} Thompson, D.J., et al., 1993, Astrophys. J. Supp., 86, 629.

\bibitem{extragal} Sreekumar, P., et al., 1998, Astrophys. J., 494, 523.

\bibitem{stefano} Vercellone, S., et al., 2001, in {\it
Probing the Physics of Active Galactic Nuclei by Multiwavelength
Monitoring}, NASA-GSFC Greenbelt 2000, eds. B.M. Peterson, R.S.
Polidan \& R.W. Pogge, ASP Conf. Series 224, p.483-490.

\bibitem{galcent} Thompson, D.J., et al., 1993, Astrophys. J. Supp., 86, 629.

\bibitem{comptel} Schoenfelder V., et al., 1993, Astrophys. J. Supp., 86, 657.

\end{thebibliography}
\end{document}